\documentclass[a4paper,10pt,twoside]{cpc-hepnp}

\usepackage{multicol}
\usepackage{graphicx}
\usepackage{booktabs}
\usepackage{amssymb,bm,mathrsfs,bbm,amscd}
\usepackage[tbtags]{amsmath}
\usepackage{lastpage}
\usepackage{CJK}
\usepackage{url}
\usepackage{threeparttable}

\begin{document}
\begin{CJK*}{GBK}{song}

\fancyhead[c]{\small Chinese Physics C~~~Vol. 37, No. 1 (2013)010201} \fancyfoot[C]{\small 010201-\thepage}

\footnotetext[0]{Received 14 March 2009}

\title{The Higgs Signatures at the CEPC CDR Baseline\thanks{Supported by National Key Program for $S\&T$ Research and Development (Grant No.: 2016YFA0400400), the National Natural Science Foundation of China (Grant No.: 11675202), and the Hundred Talent programs of Chinese Academy of Science (Grant No.: Y3515540U1) }}

\author{%
Hang Zhao$^{1,2,3;1)}$\email{zhaohang@mail.ihep.ac.cn}
\quad Yong-Feng Zhu$^{1,4)}$
\quad Cheng-Dong Fu$^{1}$
\quad Dan Yu$^{1}$\\
\quad Man-Qi Ruan$^{1;2)}$\email{manqi.ruan@mail.ihep.ac.cn}
}
\maketitle

\address{%
$^1$ Institute of High Energy Physics, Chinese Academy of Sciences, Beijing 100049, China\\
$^2$ CAS Center for Excellence in Particle Physics, Beijing 100049, China\\
$^3$ Collaborative Innovation Center for Particles and Interactions, Hefei 230026, China\\
$^4$ University of Chinese Academy of Sciences, Beijing 100049, China
}

\begin{abstract}
As a Higgs factory, The CEPC (Circular Electron-Positron Collider) project aims at the precise measurement of the Higgs boson properties.
A baseline detector concept, APODIS (A PFA Oriented Detector for the HIggS factory), has been proposed for the CEPC CDR (Conceptual Design Report) study.
We explore the Higgs signatures at this baseline setup with $\nu\nu$Higgs events.
The detector performance of the charged particles, the photons, and the jets are quantified with Higgs $\to \mu\mu, \gamma\gamma$, and jet final states respectively.
The accuracy of reconstructed Higgs boson mass is comparable at different decay modes with jets in the final states.
We also analyze the Higgs $\to$ WW$^{*}$ and ZZ$^{*}$ decay modes, where a clear separation between different decay cascades is observed. 

\end{abstract}



\footnotetext[0]{\hspace*{-3mm}\raisebox{0.3ex}{$\scriptstyle\copyright$}2013
Chinese Physical Society and the Institute of High Energy Physics
of the Chinese Academy of Sciences and the Institute
of Modern Physics of the Chinese Academy of Sciences and IOP Publishing Ltd}%

\begin{multicols}{2}

\section{Introduction}
After the Higgs boson discovery~\cite{Higgs1,Higgs2} at the LHC (Large Hadron Collider), the precise measurements of the Higgs boson properties become vital for the experimental particle physics.
The CEPC, a future high energy collider project based on a 100 km circumference main ring~\cite{CEPC1}, is therefore proposed. 
Operating at 240 GeV center of mass energy, the CEPC has an instant luminosity of $\sim$3$\times10^{34} cm^{-2}s^{-1}$ and can deliver 10$^{6}$ Higgs bosons~\cite{CEPC2}.
The CEPC can determine the absolute Higgs boson couplings to the relative accuracy of  0.1\% - 1\%, roughly one order of magnitude superior to the Higgs signal strength measurements at the HL-LHC~\cite{HLLHC1,HLLHC2}.

At the CEPC, the Standard Model (SM) Higgs bosons are produced mainly through the Higgsstrahlung process ( $e^{+}e^{-}\to ZH$) and the vector boson fusion processes (the Z fusion process $e^{+}e^{-}\to e^{+}e^{-}H$, and the W fusion process $e^{+}e^{-}\to \nu\bar{\nu}H$), see Figure~\ref{fig1}.
The corresponding cross sections with a 125 GeV SM Higgs boson using non-polarized beam at different center of mass energy is shown in Figure~\ref{fig2}.
At the CEPC, roughly a quarter of the Higgs boson is generated in association with a pair of neutrinos ($\nu\nu$H), including both the W fusion events and the ZH events with Z decays to $\nu\nu$.
The Higgs boson is responsible for almost all the detector signals in these $\nu\nu$H events, providing benchmark samples for the CEPC detector performance study.

\begin{center}
\includegraphics[width=10cm, height=4.5cm]{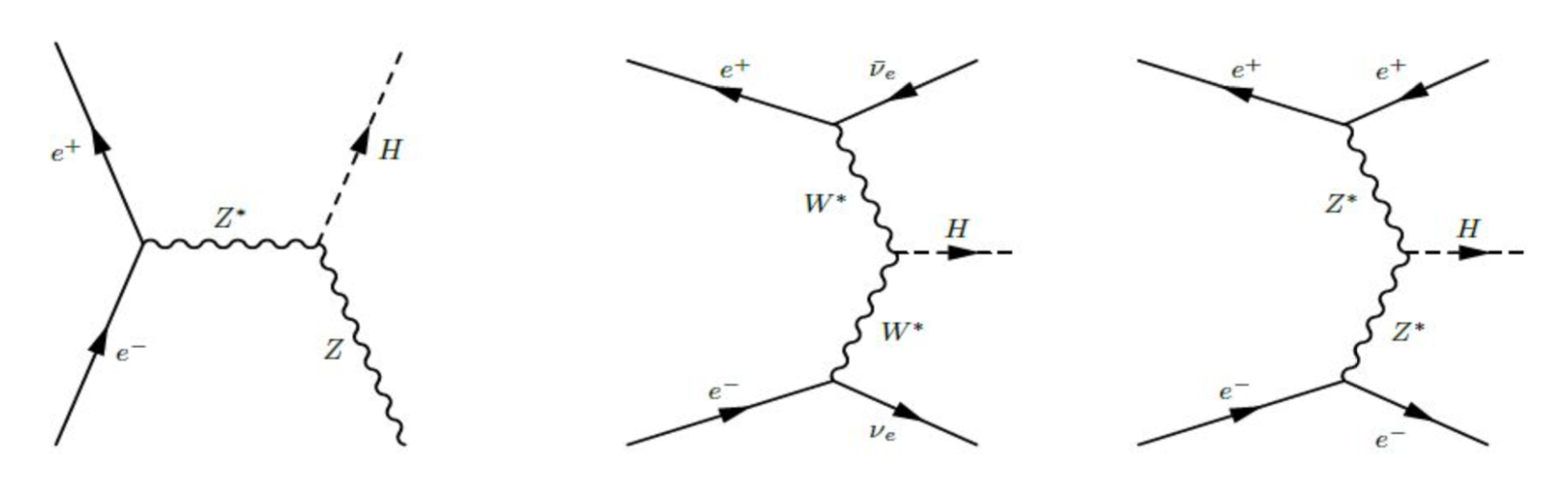}
\figcaption{\label{fig1} The Higgs boson production processes at the CEPC. }
\end{center}

This manuscript presents the performance analysis of the CEPC baseline detector geometry APODIS (a.k.a. the CEPC\_v4)~\cite{APODIS}.
Using full simulated $\nu\nu$H samples, we analyze a set of Higgs signal distributions that covers all the major SM Higgs decay modes. 
In section 2, we introduce the baseline detector and the software tools. 
Section 3 shows the reconstruction results of the Higgs signals. 
A conclusion is given in section 4.

\begin{center}
\includegraphics[width=8cm]{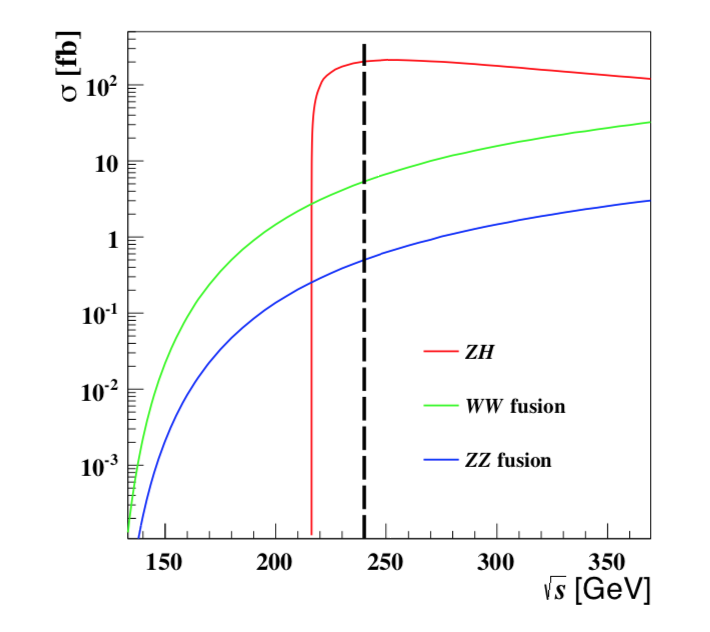}
\figcaption{\label{fig2} The Higgs boson production cross section as a function of center of mass energy. }
\end{center}

\section{Baseline detector and CEPC software chain}
The APODIS is optimized from the CEPC\_v1~\cite{CEPCv1}, the reference detector design for the CEPC Pre-CDR~\cite{CEPC1} study. 
Comparing to CEPC\_v1, APODIS maintains the same level of performance for the Higgs signals (the lepton identification and the reconstruction of photon and jets), and enhances significantly the performance on the charged Kaon identification~\cite{APODIS}. 
Meanwhile, APODIS has significantly reduced the number of readout channels, the total weight, and the solenoid B-field.

In terms of the software, a full simulation-reconstruction toolkit has been established and optimized for the APODIS geometry. 
The information flow consists of three basic modules: the Generator, the Simulation, and the Reconstruction, see Figure~\ref{fig3}. 
For the Generator, Whizard~\cite{WhizardRef} is used to generate final state particles for given physics process.
The samples are then fully simulated using the MokkaPlus~\cite{MokkaP,MokkaRef} and reconstructed using Arbor~\cite{ArborRef} as the core Particle Flow~\cite{PFARef} reconstruction. 

We simulate and reconstruct the $\nu\nu$Higgs, Higgs$\to \gamma\gamma$, $\mu\mu$, gg, bb, cc, WW$^{*}$, ZZ$^{*}$ samples, each with $\sim$50000 events statistic.
Most of the Standard Model Higgs boson decay modes are included except the Higgs $\to \tau\tau$, which has been extensively studied in the reference~\cite{Htau}.

\begin{center}
\includegraphics[width=10cm]{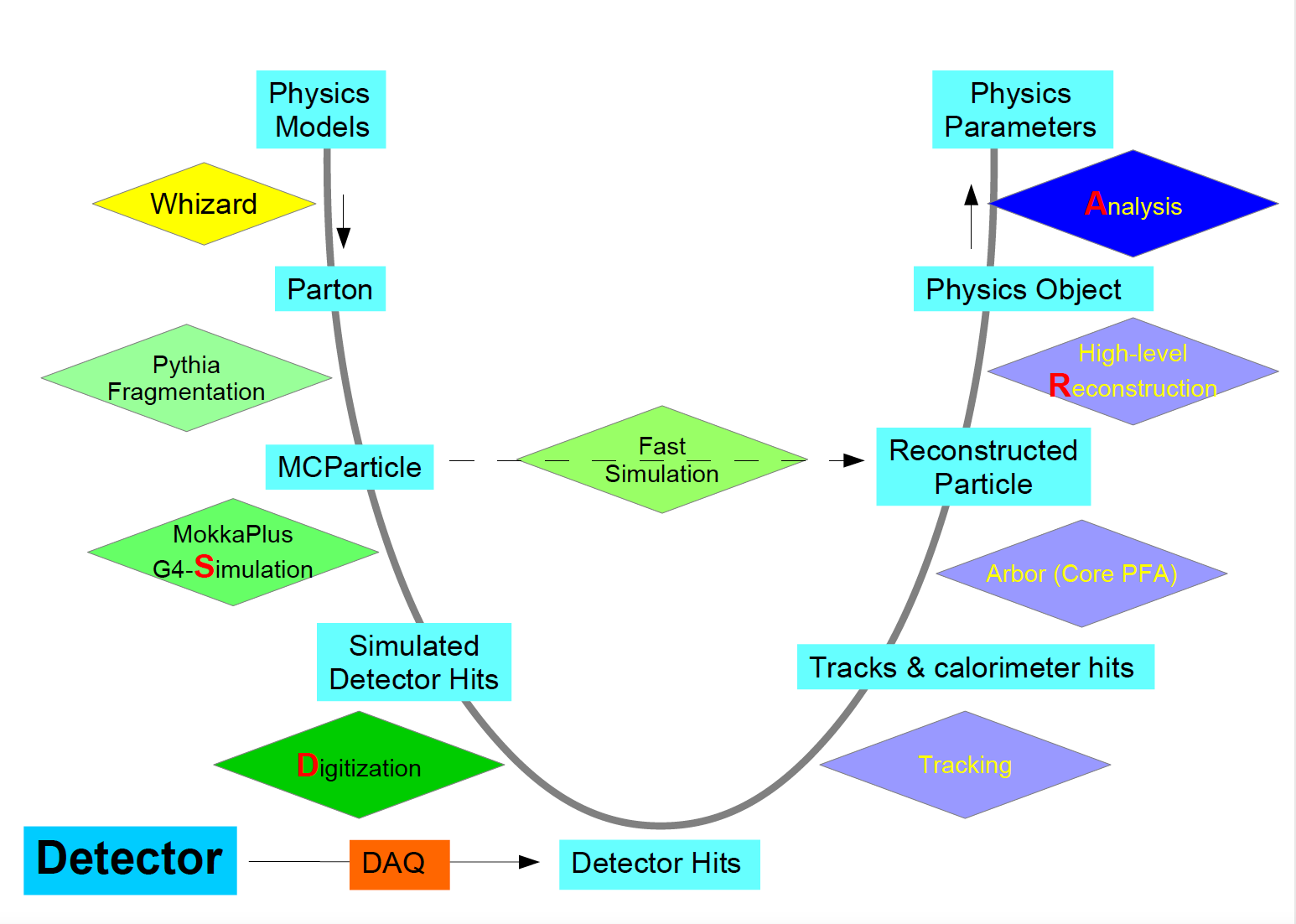}
\figcaption{\label{fig3} The information flow of the CEPC software chain. }
\end{center}

\section{Benchmark distributions of the Higgs signals at $\nu\nu$H events}
The different decay modes can be classified into three classes according to their dependence to the detector performance. 
The invariant mass reconstructed from Higgs$\rightarrow \mu\mu$ events represents the tracking reconstruction performance, and that from Higgs$\rightarrow \gamma\gamma$ represents the performance of photon reconstruction.
The other events mostly depend on the jet reconstruction performance.
The muons and photons are single particle level physics objects that can be directly identified from the reconstructed particles.
For the other decay mode events (Higgs$\rightarrow$ bb, cc, gg, WW$^{*}$, ZZ$^{*}$), we looked into the total invariant mass of the events. 
In the full reconstructed events, the total invariant mass distributions not only represent the detector performance but also include other physics effects:
\begin{itemize}
\item The ISR (initial state radiation) photons.
\item The neutrinos generated from the Higgs bosons.
\item The direction of the jets at the Higgs$\to$di-jets events, due to the acceptance of the detector.
\end{itemize}
Figure~\ref{fig4} shows the correlation between reconstructed Higgs boson mass and the sum of the transverse momentum of the ISR photons (Pt\_ISR), the sum of the transverse momentum of the neutrinos generated from Higgs boson (Pt\_neutrino), and the minimum angle between the jets and the beam pipe ($|$Cos(Theta\_Jet)$|$), at the Higgs $\to$ di-gluon events.
Clearly a strong correlation is observed when these effects are significant.
In order to disentangle these effects from the detector performance at jet reconstruction, a monte calo truth level event selection is applied to the events with jets in the final states. 
The standard cleaning procedure is set up as the Pt\_ISR $<$ 1 GeV, the Pt\_neutrino $<$ 1 GeV and the $|$Cos(Theta\_Jet)$|$ $<$ 0.85, with the selection efficiency shown in Table~\ref{tab1}. 

\end{multicols}
\begin{center}
\includegraphics[width=5cm]{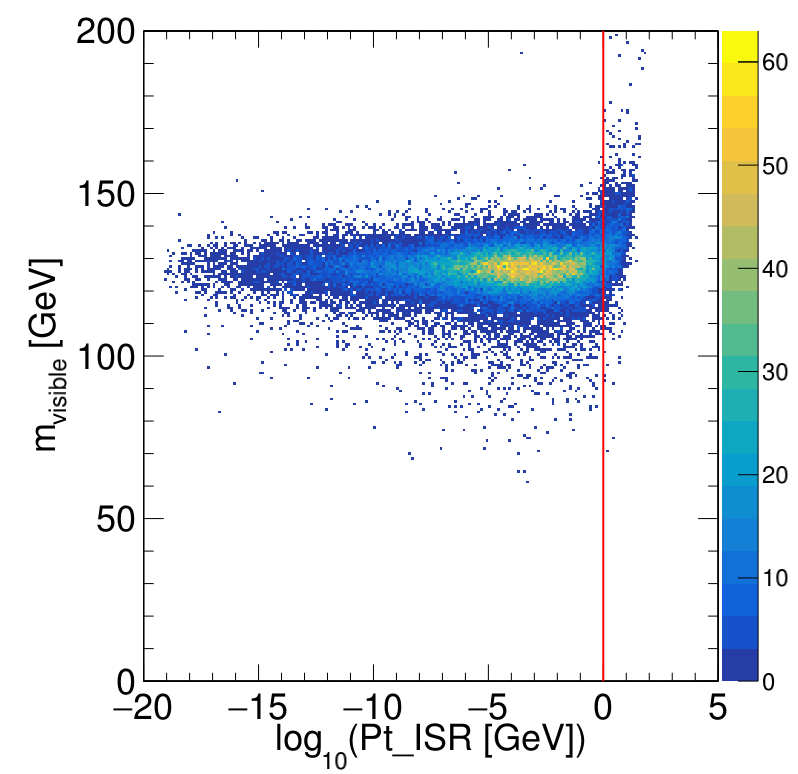}
\qquad
\includegraphics[width=5cm]{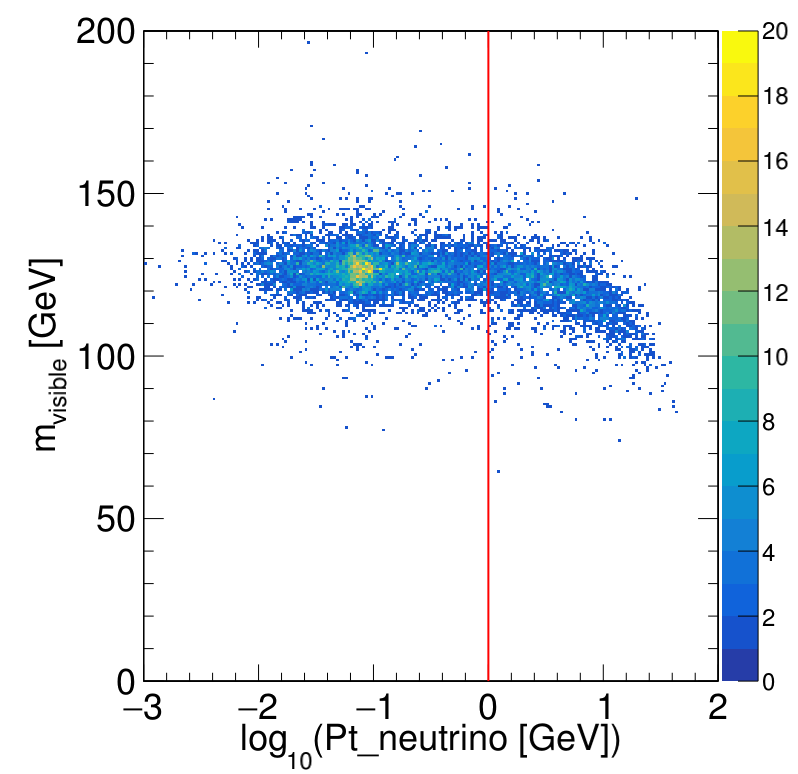}
\qquad
\includegraphics[width=5cm]{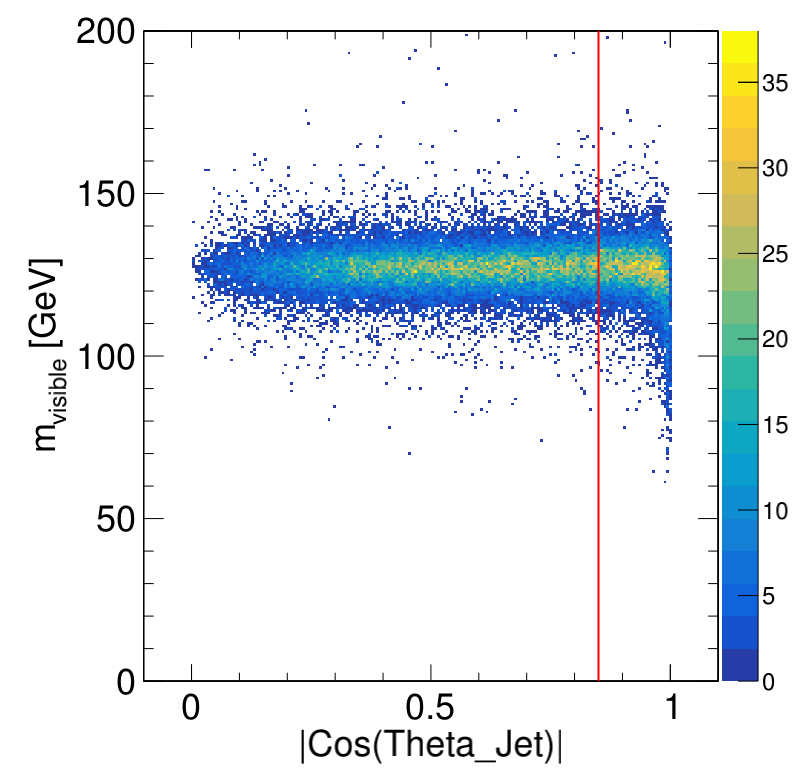}
\figcaption{\label{fig4} The correlation between reconstructed Higgs boson mass and the sum of the transverse momentum of the ISR photons (Pt\_ISR), the sum of the transverse momentum of the neutrinos generated from Higgs boson (Pt\_neutrino), and the the minimum angle between the jets and the beam pipe ($|$Cos(Theta\_Jet)$|$), at the Higgs $\to$ di-gluon events. }
\end{center}
\begin{multicols}{2}
 
\end{multicols}
\begin{center}
\tabcaption{ \label{tab1}  Event selection efficiency for Higgs boson exclusive decay at CEPC with $\sqrt{s}=240$ GeV.}
\footnotesize
\begin{tabular*}{170mm}{@{\extracolsep{\fill}}cccccccc}
\toprule	&	gg	&	bb		&	cc		&	WW$^{*}	$	& ZZ$^{*}	$\\
\hline
Total Events &	49092			&	48931 	&	50000	&	50000			&	49123 \\
\hline
Pt\_ISR $<$ 1GeV      &   $95.15\%$   & $95.37\%$   & $95.30\%$   & $95.16\%$           &  $95.24\%$ \\
\hline
Pt\_neutrino $<$ 1GeV    &    $89.33\%$  & $39.04\%$ & $66.36\%$  & $37.46\%$        & $41.39\%$                  \\
 \hline
$|$Cos(Theta\_Jet)$|$ $<$ 0.85       &    $67.30\%$ &$28.65\%$   & $49.31\%$ &  -        &  -    \\
\bottomrule
\end{tabular*}
\end{center}

\begin{multicols}{2}

\begin{center}
\includegraphics[width=7cm]{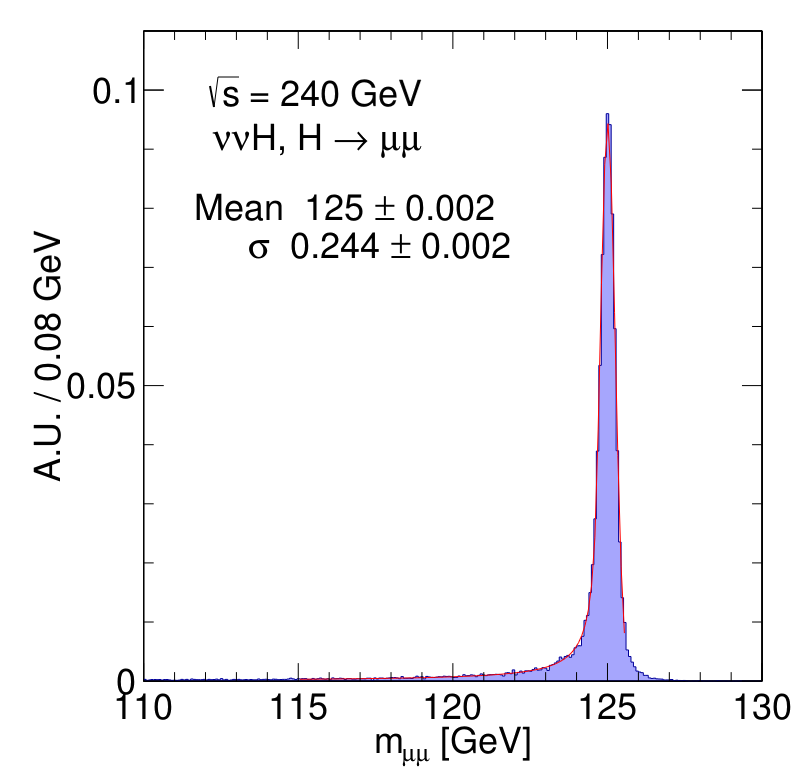}
\figcaption{\label{fig5} The distribution of reconstructed Higgs boson mass at Higgs $\to \mu\mu$ events, fitted to a crystal-ball function.}
\end{center}

\begin{center}
\includegraphics[width=7cm]{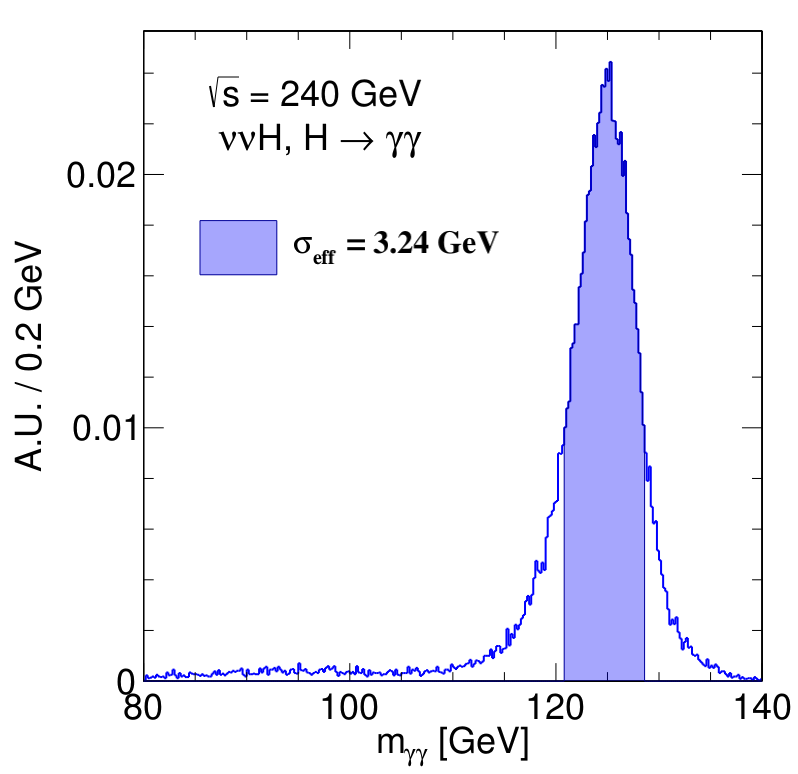}
\figcaption{\label{fig6} The distribution of reconstructed Higgs boson mass at Higgs $\to \gamma\gamma$ events, parameterized by the half-width of the narrowest interval containing 68.3\% of the distribution ($\sigma_{eff}$).}
\end{center}

\subsection{Higgs $\rightarrow \mu\mu$}
The Higgs boson decay into $\mu^{+}\mu^{-}$ is a rare process with a branching ratio of 0.022\% for an 125 GeV SM Higgs boson~\cite{HBR}.
From the reconstructed particles, we select a pair of muons with opposite charge and both have energy higher than 20 GeV. 
The invariant mass of this pair of muons is reconstructed as the Higgs boson mass.
The distribution fitted to a crystal-ball function is shown in Figure~\ref{fig5}.
The tracking performance is characterized by the resolution of the Higgs boson mass, which reaches 0.20\% ($\sigma$/Mean) in this benchmark channel.

\subsection{Higgs$ \rightarrow \gamma \gamma$}
The performance of the ECAL (electromagnetic calorimeter) is characterized by the reconstruction of the Higgs $\to \gamma\gamma$ events. 
We select two most energetic photons with energy higher than 10 GeV and calculate their invariant mass as the Higgs boson mass, see Figure~\ref{fig6}.
The tail at the left side is caused by the geometry defects and the material before the calorimeter.
The width of the distribution is parameterized by the half-width of the narrowest interval containing 68.3\% of the distribution, $\sigma_{eff}$~\cite{CMSPhotonRec}.
The $\sigma_{eff}$ of the distribution is 3.24 GeV and the resolution ($\sigma$/Mean) is 2.59\%.

At a simplified geometry free of the geometry defects, the reconstructed Higgs boson mass resolution can reach 1.64\%~\cite{SimplifiedGeo}.
This significant degradation at the APODIS indicates the geometry based correction of the photon energy reconstruction is mandatory. 

\end{multicols}
\begin{center}
\includegraphics[width=5cm]{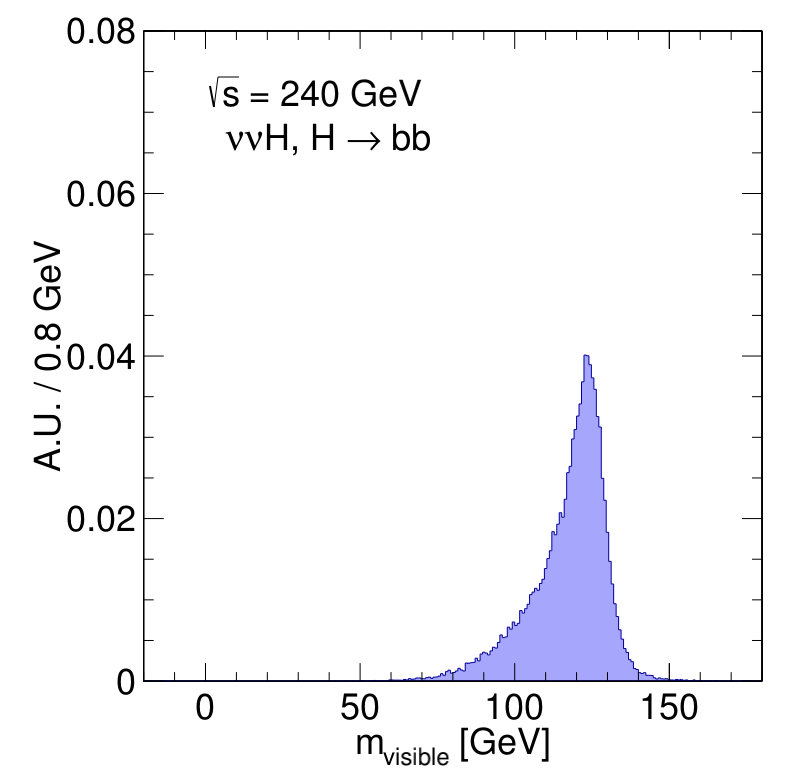}
\qquad
\includegraphics[width=5cm]{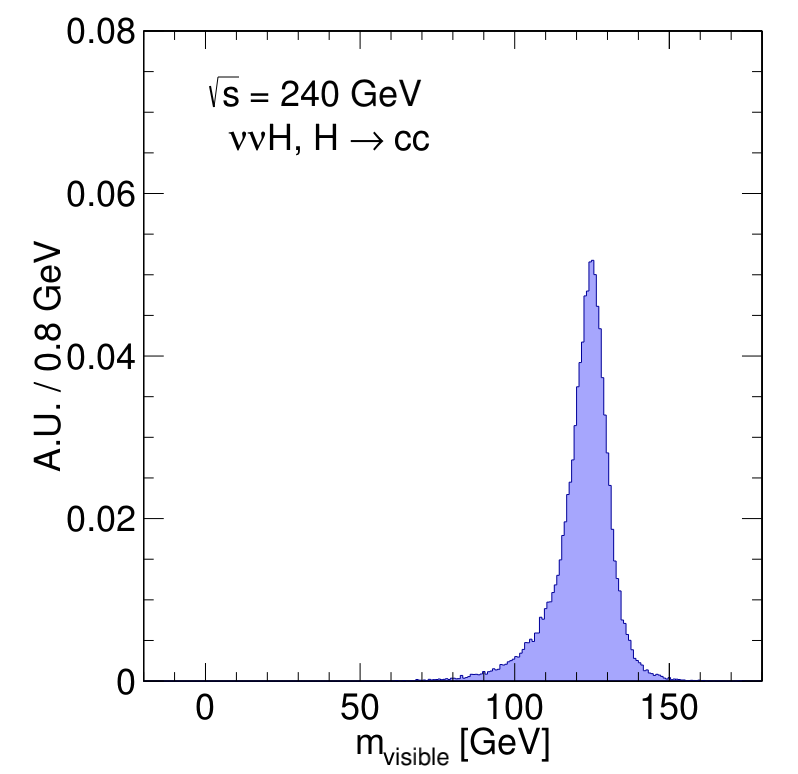}
\qquad
\includegraphics[width=5cm]{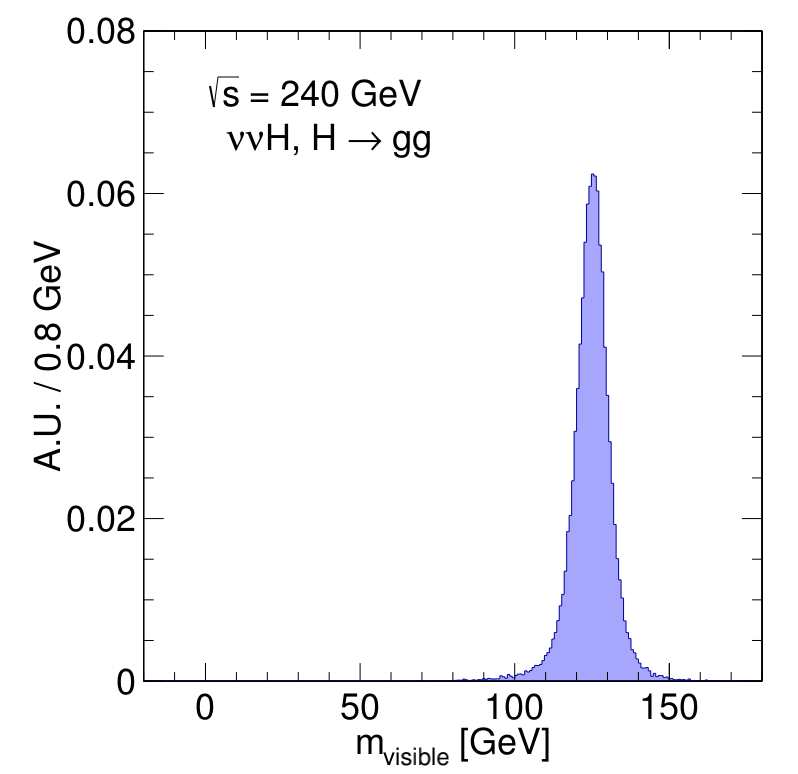}
\figcaption{\label{fig7} The distributions of reconstructed total visible invariant mass at the Higgs $\to$ bb, cc, gg events before the event cleaning. }
\end{center}

\begin{center}
\includegraphics[width=5cm]{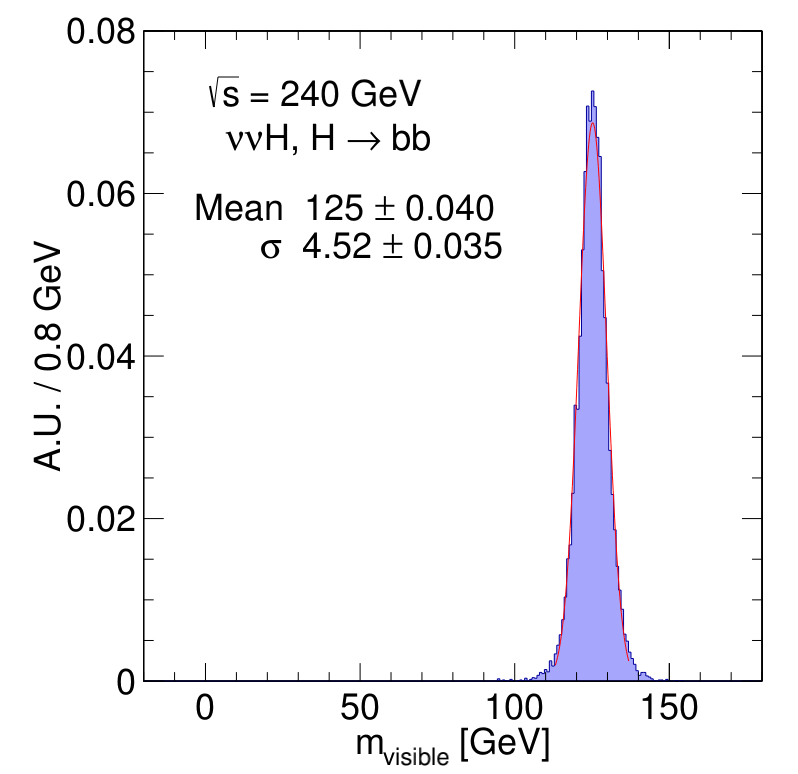}
\qquad
\includegraphics[width=5cm]{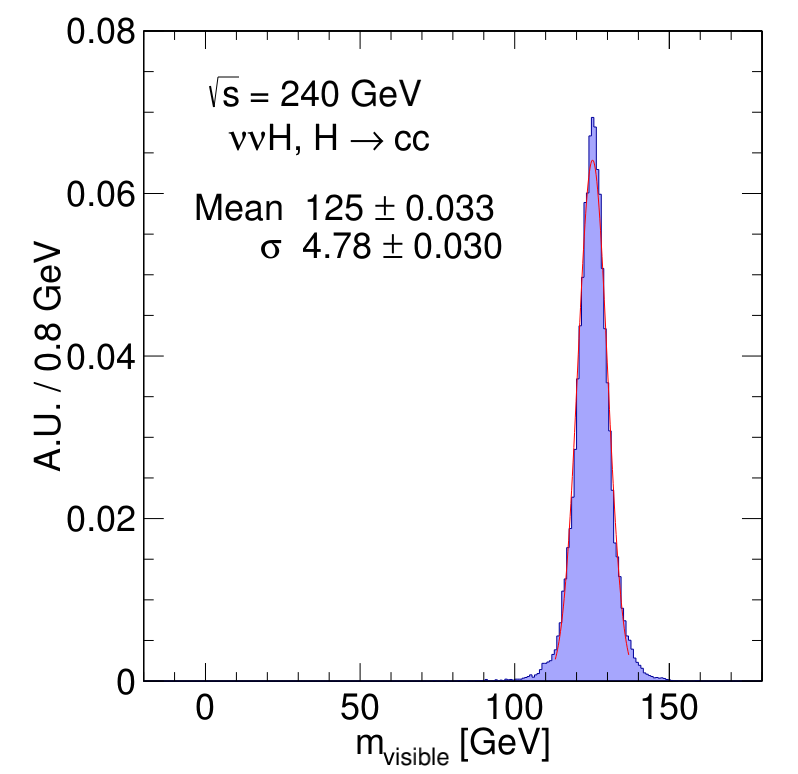}
\qquad
\includegraphics[width=5cm]{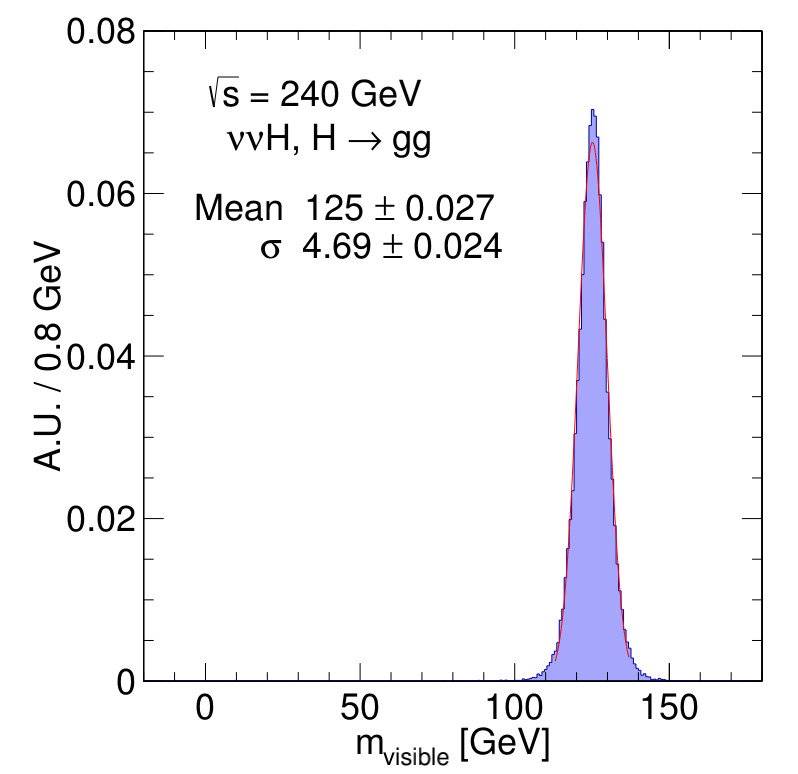}
\figcaption{\label{fig8} The distributions of reconstructed total visible invariant mass at the Higgs $\to$ bb, cc, gg events after the event cleaning and fitted with Gaussian function. }
\end{center}
\begin{multicols}{2}

\subsection{Higgs$ \rightarrow$ bb, cc, gg}
Roughly 70\% of the 125 GeV SM Higgs bosons decay into a pair of jets (bb, cc, and gg).
For these Higgs $\to$ di-jets events, we collect all the visible final state particles and calculate their invariant mass. 
The distributions of reconstructed Higgs boson mass before event selection are shown in Figure~\ref{fig7} and the Figure~\ref{fig8} shows the results after applying the event selection.
After the cleaning, the resolutions of Higgs boson mass at different Higgs $\to$ di-jets events are almost consistent, which are 3.63\%(bb), 3.82\%(cc), and 3.75\%(gg).
 
\subsection{Higgs$ \rightarrow WW^{*}$}
The 125 GeV SM Higgs boson has a probability of 21.4\% to decay into a pair of W bosons, making this measurement a sensitive probe to the New Physics. 
Limited by the Higgs boson mass, one of the W boson is off-shell (W$^{*}$). 

The reconstructed total visible invariant mass distribution is shown in Figure~\ref{fig9}.
Depending on the decay modes of W and W$^{*}$ (leptonic or hadronic), the total invariant mass distribution is decomposed into different sub-distributions. 
The main peak at 125 GeV corresponding to the events that both W and W$^{*}$ decay into two quarks.
The other two stacks corresponding to the events that W or W$^{*}$ decay into a pair of lepton and neutrino.

\begin{center}
\includegraphics[width=7cm]{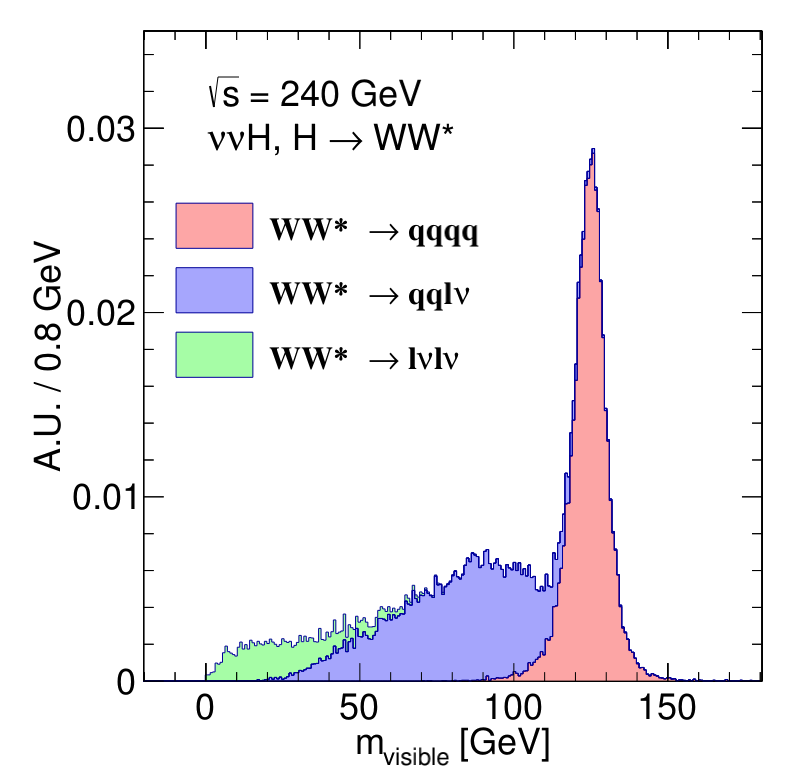}
\figcaption{\label{fig9} The distribution of reconstructed total visible invariant mass at Higgs $\to WW^{*}$ events. Depending on the decay modes of W and W$^{*}$ (leptonic or hadronic), the total invariant mass distribution is decomposed into different sub-distributions.}
\end{center}
 
The event selection procedure is also applied.
After the event cleaning, the events that W or W$^{*}$ decay into leptons and neutrinos are excluded, as shown in Figure~\ref{fig10}.
This cleaned invariant mass distribution can fit to a Gaussian function with $\sigma$=4.77.
The fit result is comparable with the Higgs $\rightarrow$ di-jets events.

\begin{center}
\includegraphics[width=7cm]{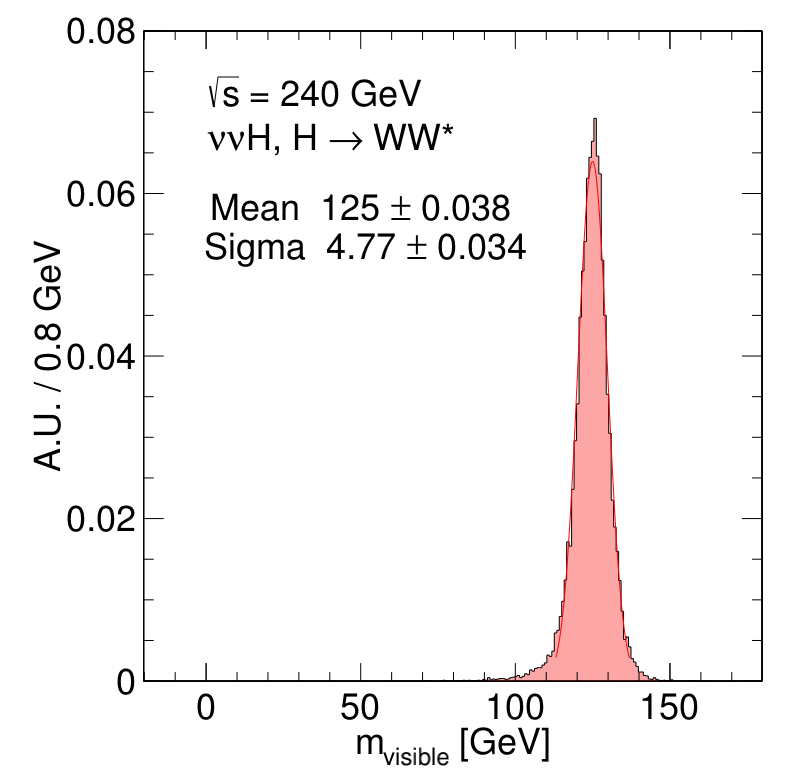}
\figcaption{\label{fig10} The distribution of reconstructed total visible invariant mass at Higgs $\to WW^{*}$ events after the event cleaning, fitted to a Gaussian function.}
\end{center}

\subsection{Higgs$ \rightarrow ZZ^{*}$}
About 2.6\% of the 125 GeV SM Higgs boson decay into $ZZ^{*}$.
Similar to Higgs$ \rightarrow WW^{*}$ channel, one of the Z boson is off-shell. 
Most of the Z bosons then decay into $q\bar{q}$ ($\sim$70\%), $l\bar{l}$ ($\sim$10\%) or $\nu\nu$ ($\sim$20\%).

Figure~\ref{fig11} shows the reconstructed total invariant mass.
With the monte calo truth information, the events are classified depending on the decay modes of Z and Z$^{*}$ (visible or invisible).
There are four peaks in the distribution.
The peak at zero is corresponding to the events that both Z and Z$^{*}$ decay into neutrinos, which contains about 4\% of all the events. 
The main peak at the expected Higgs boson mass ($\sim$125 GeV) is corresponding to the events with all the final state particles are visible.
The other two peaks are corresponding to the conjugation case that $Z\to$ visible, $Z^{*} \to$ invisible and $Z^{*} \to$ visible, $Z\to$ invisible. 

\begin{center}
\includegraphics[width=7cm]{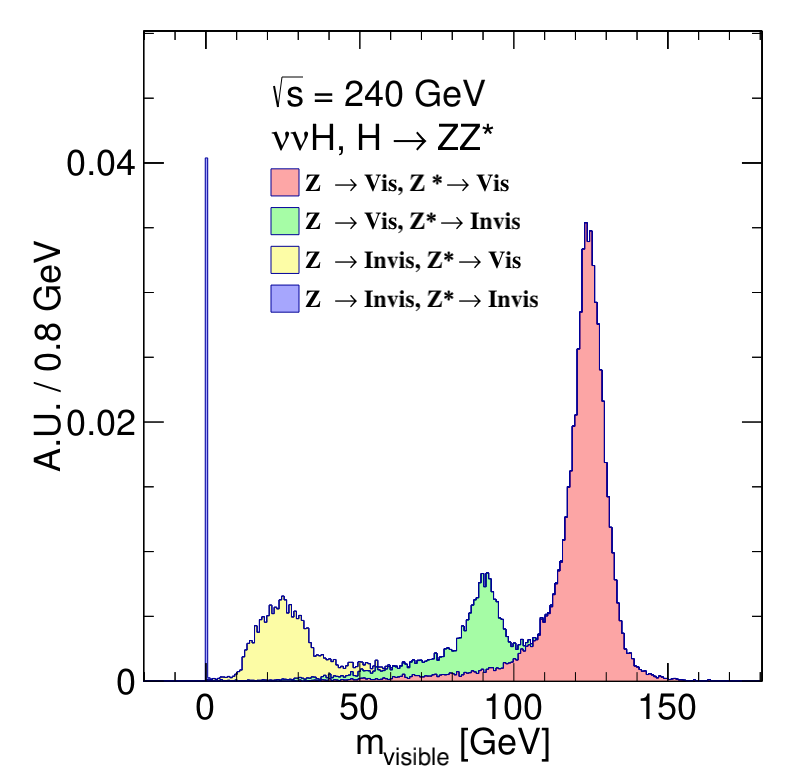}
\figcaption{\label{fig11} The distribution of reconstructed total invariant mass at Higgs $\to ZZ^{*}$ events, classified depending on the decay modes of Z and Z$^{*}$ (visible or invisible).}
\end{center}

By using the event selection procedure, the events Z or $Z^{*}$ decay into neutrinos are excluded, as shown in Figure~\ref{fig12}.
Determined by the Z decay modes, most of the remaining events contain 4 or 2 quarks as final state particles.
This cleaned invariant mass distribution can fit to a Gaussian function with $\sigma$=4.68.
The resolution is also comparable with the Higgs $\to$ di-jets events. 

\begin{center}
\includegraphics[width=7cm]{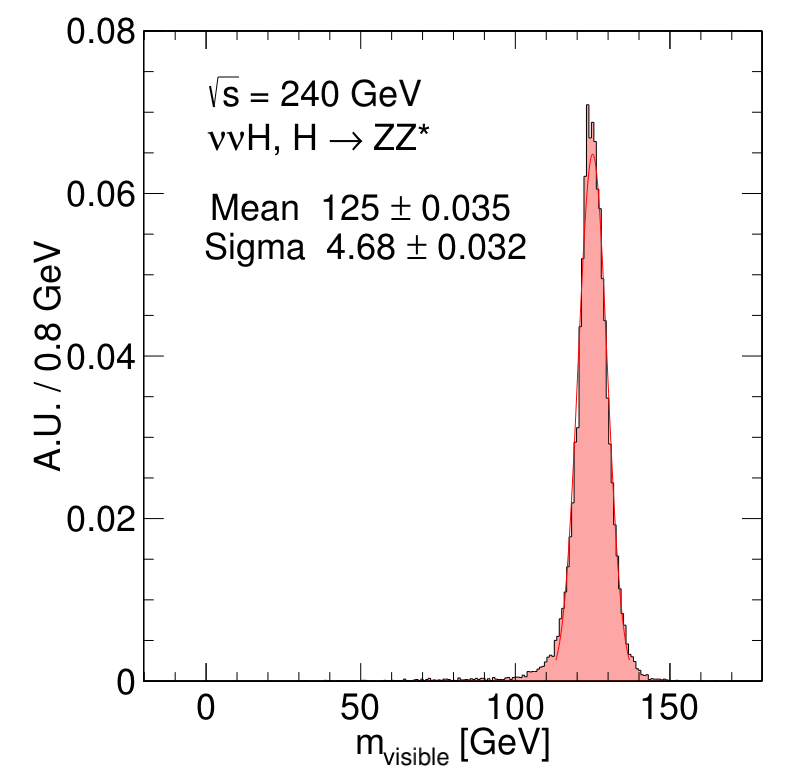}
\figcaption{\label{fig12} The distribution of reconstructed total visible invariant mass at Higgs $\to ZZ^{*}$ events after the event selection, fitted to a Gaussian function.}
\end{center}

\section{Conclusion}
Based on the APODIS detector design, we characterize the Higgs signatures at the $e^{+}e^{-}  \to \nu\nu Higgs$ events with Higgs decaying to $\gamma\gamma$, $\mu\mu$, gg, bb, cc, $WW^{*}$ and $ZZ^{*}$.
With the  Higgs $\to \mu\mu, \gamma\gamma$, and jet final states events, we quantify the detector performance, as shown in Table~\ref{tab2}.
Comparing to the results at LHC, the reconstruction accuracy at Higgs $\to \mu\mu$ events is improved by about one magnitude, and that at Higgs $\to$ di-jets events is improved by about 3 times.
The resolution at Higgs $\to \gamma\gamma$ events degrades by roughly 60\%, limited by the absence of geometry based correction and fine-tuned calibration, and the sampling fraction of ECAL. 

To describe the jet reconstruction performance, a standard event cleaning procedure has been designed.
After the event cleaning, the accuracy of the reconstructed Higgs boson mass at different Higgs decay modes with jets as final state particles are comparable, as shown in Table~\ref{tab3}.

For the Higgs $\to$ WW$^{*}$ and ZZ$^{*}$ events, the total invariant mass distribution is composed of multiple components depending on the decay modes of W and Z bosons.
For the WW$^{*}$ events, the classification is based on the leptonic or hadronic decay mode of W and W$^{*}$.
For the ZZ$^{*}$ events, the reconstruction result is sensitive to the visible or invisible decay mode of Z and Z$^{*}$.
The distribution of H $\rightarrow$ ZZ$^{*}$ is clearly separated with four peaks corresponding to the Z and Z$^{*}$ decay modes.
The standard cleaning procedure could efficiently veto the events with significant neutrinos generated from the Higgs boson cascade decay.
After the event cleaning, the Higgs boson mass resolutions ($\sigma$/Mean) at Higgs $\to$ WW$^{*}$ and ZZ$^{*}$ events are comparable with Higgs $\rightarrow$ di-jets events, see Table~\ref{tab3}.

\acknowledgments{
We would like to thank Gang Li and Xin Mo for the physics event generator files.
}

\end{multicols}
\begin{center}
\begin{threeparttable}[b]
\tabcaption{ \label{tab2}  Benchmark resolutions ($\sigma/Mean$) of reconstructed Higgs boson mass, comparing to LHC results.}
\footnotesize
\begin{tabular*}{170mm}{@{\extracolsep{\fill}}cccc}
\toprule        & Higgs$\to \mu\mu$    &   Higgs$ \to \gamma\gamma$.      & Higgs$\to$bb   \\
\hline
CEPC (APODIS)     &      0.20\%            &     2.59\%\tnote{1}         &          3.63\%              \\
LHC (CMS, ATLAS) &      $\sim$2\%~\cite{LHCHmumu1, LHCHmumu2}    &     $\sim$1.5\%~\cite{LHCHaa1, LHCHaa2}         &          $\sim$10\% ~\cite{LHCHbb1, LHCHbb2}             \\
\bottomrule
\end{tabular*}
\begin{tablenotes}
\item[1] primary result without geometry based correction and fine-tuned calibration.
\end{tablenotes}
\end{threeparttable}

\end{center}
\begin{multicols}{2}

\end{multicols}
\begin{center}
\tabcaption{ \label{tab3}  Higgs boson mass resolution ($\sigma/Mean$) at different decay modes with jets as final state particles, after the event cleaning.}
\footnotesize
\begin{tabular*}{170mm}{@{\extracolsep{\fill}}ccccc}
\toprule       Higgs$\to$bb       &   Higgs$\to$cc   & Higgs$\to$gg & Higgs$\to$ WW$^{*}$    & Higgs$\to$ ZZ$^{*}$       \\
\hline
       3.63\%      &     3.82\%     &        3.75\%          &        3.81\% &         3.74\% \\
\bottomrule
\end{tabular*}
\end{center}
\begin{multicols}{2}

\end{multicols}

\vspace{10mm}

\vspace{-1mm}
\centerline{\rule{80mm}{0.1pt}}
\vspace{2mm}

\begin{multicols}{2}

\end{multicols}

\clearpage

\end{CJK*}

\begin{thebibliography}{90}

\vspace{3mm}

\bibitem{Higgs1}
ATLAS Collaboration, G. Aad et al., \emph{Observation of a new particle in the search for the Standard Model Higgs boson with the ATLAS detector at the LHC},\emph{Phys. Lett. B} {\bf716}(2012) 1-29, arXiv:1207.7214 [hep-ex].
\bibitem{Higgs2}
CMS Collaboration, S. Chatrchyan et al., \emph{Observation of a new boson at a mass of 125 GeV with the CMS experiment at the LHC},\emph{Phys. Lett. B} {\bf716}(2012) 30-61, arXiv:1207.7235 [hep-ex].
\bibitem{CEPC1}
The CEPC-SPPC Study Group, \emph{CEPC-SPPC Preliminary Conceptual Design Report}, \url{http://cepc.ihep.ac.cn/preCDR/main_preCDR.pdf}
\bibitem{CEPC2}
Xinchou Lou, \emph{Circular Electron-Positron Collider - Challenges and Opportunities}, presentation at IAS HEP conference, 2018, \url{http://ias.ust.hk/program/shared_doc/2018/201801hep/conf/talks/HEP_20180123_0945_Xinchou_Lou.pdf}
\bibitem{HLLHC1}
CMS Collaboration, \emph{Projected Performance of an Upgraded CMS Detector at the LHC and HL-LHC: Contribution to the Snowmass Process}, 2013, ArXiv: 1307.7135
\bibitem{HLLHC2}
ATLAS Collaboration, \emph{Physics at a High-Luminosity LHC with ATLAS}, 2013, ArXiv: 1307.7292
\bibitem{APODIS}
Manqi Ruan, \emph{CEPC Baseline detector concept}, presentation at the Workshop on the CEPC-EU edition, 2018, \url{https://agenda.infn.it/getFile.py/access?contribId=23&sessionId=23&resId=0&materialId=slides&confId=14816}
\bibitem{CEPCv1}
M. Ruan, \emph{Simulation, reconstruction and software at the CEPC, presentation at CEPC workshop}, 2016, \url{http://indico.ihep.ac.cn/event/5277/session/14/contribution/67/material/slides/0.pdf}
\bibitem{CEPCSoftWeb}
\emph{CEPC software website}, \url{http://cepcsoft.ihep.ac.cn/}
\bibitem{WhizardRef}
W. Kilian et al., \emph{Simulating Multi-Particle Processes at LHC and ILC}, Eur.Phys.J.C71 (2011) 1742, arXiv: 0708.4233 [hep-ph]
\bibitem{MokkaRef}
P. Mora de Freitas and H. Videau, \emph{Detector simulation with MOKKA / GEANT4: Present and future}
\bibitem{MokkaP}
Source code of MokkaPlus, \url{http://cepcgit.ihep.ac.cn/cepcsoft/MokkaC}.
\bibitem{ArborRef}
M. Ruan and H. Videau, Arbor, \emph{a new approach of the Particle Flow Algorithm}, arxiv:1403.4784 [physics.ins-det]
\bibitem{PFARef}
M. Thomson, \emph{Particle Flow Calorimetry and the PandoraPFA Algorithm}, \emph{Nucl.Instrum.Meth} {\bf A611}(2009) 25,40. arxiv:0907.3577[physics.ins-det]
\bibitem{Htau}
D. Yu, \emph{Reconstruction of leptonic physic objects at future e+e- Higgs factory}, Ph.D Thesis, 2018
\bibitem{HBR}
C. Patrignani et al., (Particle Data Group), \emph{Review of Particle Physics}, Chinese Physics C, 40, 100001 (2016)
\bibitem{CMSPhotonRec}
CMS collaboration, \emph{Performance of photon reconstruction and identification with the CMS detector in proton-proton collisions at $\sqrt{s}$ = 8TeV}, JINST 10 (2015) P08010, arXiv:1502.02702
\bibitem{SimplifiedGeo}
H. Zhao et al., \emph{Particle flow oriented electromagnetic calorimeter optimization for the circular electron positron collider}, Journal of Instrumentation, Volume 13, March 2018, arXiv:1712.09625.
\bibitem{LHCHmumu1}
Alessandro Calandri, \emph{Prospects for Higgs physics at the HL-LHC}, presentation at IRN Terascale 2017, \url{https://indico.in2p3.fr/event/16525/contributions/58198/attachments/45869/57154/Calandri_14122017.pdf}
\bibitem{LHCHmumu2}
ATLAS Collaboration, \emph{Letter of Intent for the Phase-II Upgrade of the ATLAS Experiment}, CERN-LHCC-2012-022, LHCC-I-023, \url{https://cds.cern.ch/record/1502664/files/LHCC-I-023.pdf}
\bibitem{LHCHaa1}
CMS collaboration, \emph{Measurements of Higgs boson properties in the diphoton decay channel in proton-proton collisions at $\sqrt{s}$ = 13 TeV}, CMS-HIG-16-040, CERN-EP-2018-060, arXiv:1804.02716
\bibitem{LHCHaa2}
ATLAS Collaboration, \emph{Measurements of Higgs boson properties in the diphoton decay channel with 36 fb$^{1}$ of pp collision data at $\sqrt{s}$ =13 TeV with the ATLAS detector}, CERN-EP-2017-288, arXiv:1802.04146 
\bibitem{LHCHbb1}
ATLAS Collaboration, \emph{Evidence for the H $\to$ $b\bar{b}$ decay with the ATLAS detector}, CERN-EP-2017-175, JHEP 12 (2017) 024, arXiv:1708.03299 
\bibitem{LHCHbb2}
Manuel Daniel Proissl, \emph{Dijet Invariant Mass Studies in the Higgs boson H $\to$ $b\bar{b}$ resonance search in association with a W/Z boson using the ATLAS detector}, 2015, \url{http://hdl.handle.net/1842/10516}

\end{thebibliography}
\end{document}